\documentclass[doublecol]{epl2}
\usepackage{subfigure}
\usepackage{epsfig}
\title{Coexisting
stochastic and coherence resonance in a mean-field dynamo model for 
Earth's magnetic field reversals}
\shorttitle{Stochastic and coherence resonance in a dynamo model for 
Earth's magnetic field reversals} 
\author{M. Fischer \and F. Stefani \and  G. Gerbeth}
\shortauthor{M. Fischer \etal}
\institute{Forschungszentrum Dresden-Rossendorf, P.O. Box 510119, D-01314 Dresden
}
\pacs{91.25.Mf}{Magnetic field reversals: process and timescale}
\pacs{91.25.Cw}{Origins and models of magnetic fields; dynamo theories}
\pacs{05.40.Ca}{Noise}
\abstract{Using a spherical symmetric mean field $\alpha^2$-dynamo
model for Earth's magnetic field reversals, we show the
coexistence of the noise-induced phenomena coherence resonance 
and stochastic resonance. Stochastic resonance has been recently
invoked to explain the 100 kyr periodicity in the distribution of
the residence time between reversals.
The comparison of the resulting residence time distribution with
the paleomagnetic one allows for some estimate of
the effective  diffusion time of the Earth's core which may be
100 kyr or slightly below rather than
200 kyr as it would result from the molecular resistivity.}
\begin{document}
\maketitle
There is ample paleomagnetic evidence that the
Earth's magnetic field has undergone polarity changes
many 
times \cite{MERRIL}. 
Averaged over the last few million years the
mean rate of reversals is approximately 4-5 per Myr, 
although the last reversal occurred approximately
780000 years ago.
At least two so-called superchrons have been identified
as periods of some tens of millions of years with
no reversal at all. One
of the most intriguing features of reversals is
their pronounced asymmetry  with the initial
decay of the dipole being
much slower than the subsequent
recreation of the dipole with opposite
polarity \cite{VALET2005}.
Despite the general irregularity of the reversal time series,
there are (at least)
two features pointing to some sort of ordering.
The first one is the clustering property of
reversals \cite{CARBONE,LUCA}, the second one is
the appearance of a 100 kyr periodicity in the
distribution of the residence times
between reversals \cite{CONSOL}.

Computer simulations of
the geodynamo in
general, and of reversals in particular
\cite{WICHTOLSON},
have matured much since the first
fully coupled  3D simulation of a reversal
by Glatzmaier and Roberts in 1995 \cite{GLARO}.
However, the severe problem
that dynamo simulations have to work
in parameter regions which are far away from
the real values of the Earth's core,
will remain for a long time.
Hence it is
certainly useful to understand
the essential features of reversals in
the framework of simpler models.

With view on the recent successes of
liquid sodium dynamo
experiments \cite{RMP,MOMOMO}
simple reversal models may also help in
tailoring  future dynamo facilities to
make them prone to reversals. As a matter of fact,
reversals were observed recently \cite{BERHANU} in a
special version of the French VKS dynamo
experiment that was dominated by the use of
iron propellers with a high magnetic
permeability \cite{PETRELIS}.
However, the exponential field decay in the
initial phase of the experimentally observed
reversals seems more indicative for a  intermittent
extinguishing of the dynamo which is certainly not
equivalent to the behaviour of the geodynamo
during a reversal.
Another interesting dynamo related 
experiment showing reversals has been reported by
Bourgoin et al. \cite{BOURGOIN}.

Roughly speaking, there are two classes of simplified
dynamo models which try to explain reversals. The first
one relies on the assumption that the dominant
dipole field is somehow ''rotated'' from a given
orientation to the
reversed one via some intermediate state (or states)
\cite{HOYNG1,HOYNG2,MELBOURNE}. A typical
feature of such models,
in particular of the model of Hoyng and coworkers \cite{HOYNG1,HOYNG2},
is that they work
with an effective  conductivity of the Earth's core
which is strongly
decreased compared to the molecular value.
The necessity to introduce such a ''turbulent conductivity''
is well known for the solar dynamo \cite{OSSEN}, but
a dramatic conductivity reduction
seems hardly justified for the Earth's
core \cite{ROBERTS}.
Another drawback of the "rotation model" is that
it results in reversals with
the wrong asymmetry as it was shown recently
\cite{GAFD}.
In spite of these problems, the ''rotation model''
with a turbulent
diffusion time
of 5 kyr (instead of 200 kyr as it would result  from
the molecular resistivity)
was successful in recovering
the above mentioned 100 kyr periodicity  in the
distribution of the residence times between reversals \cite{LORITO}.
The underlying physical mechanism was identified as the
stochastic resonance (SR) \cite{GAMMAITONI}, which is
an example of noise-induced phenomena in nonlinear systems
driven by  weak periodic forcing. For the geodynamo, the
weak
periodic forcing is actually assumed to result from the
Milankovich cycle of the Earth's orbit eccentricity
\cite{YAMAZ,LIU},
although
details of the driving mechanism are hard to 
quantify from first principles.

The second class of simplified reversal models, which could be
coined "oscillation models'', relies
on a spectral peculiarity of the (in general)
non-selfadjoint dynamo operator.
The basic idea, the specific interplay between a
non-oscillatory  and
an oscillatory branch of the dominant axial dipole mode,
was expressed early by Yoshimura \cite{YOSHI}
and later exemplified
within a spherically symmetric  mean-field dynamo
model of the
$\alpha^2$-type \cite{STE1,STE2,STE3,GAFD}.
Many features of reversals, in particular the correct
time-scale, the mentioned asymmetry and
clustering property  were attributed
to the magnetic field
dynamics in the vicinity of a branching point (or
{\it exceptional point} \cite{KATO}) of the spectrum
of the dynamo operator.
Usually, this exceptional point, where two real
eigenvalues
coalesce and continue as a complex conjugated pair of
eigenvalues,  is associated with
a nearby local maximum of the
growth rate situated at a slightly lower
magnetic Reynolds number.
It is the negative slope of the growth rate
curve between this local
maximum and the exceptional point that makes
stationary
dynamos vulnerable to  noise.
Then, the instantaneous
eigenvalue is driven
towards the exceptional point and beyond into the
oscillatory branch where the sign change
happens. In this picture, reversals appear as
noise-induced relaxation oscillations.

In \cite{STE1} the model was also shown to exhibit
coherence resonance (CR) \cite{PIKOVSKY}, which is
similar to SR  but  relies on the existence of an
intrinsic frequency of the system and not on an
external periodic forcing.
Given the capability of  this model to account
for a number of
observed reversal features,  we will
ask in the present paper
how the mechanism of SR can be implemented
and what we can learn from its coexistence with
CR.  The coexistence of both resonance types 
has been demonstrated for the
Fitz-Hugh-Nagumo model with the probability density
function of the residence times showing a transition
from SR to CR
for an increasing noise strength \cite{CENT}.

After delineating our
simplified mean-field dynamo model
(more details and discussions can be found in
\cite{STE1,STE2,STE3,GAFD})
we will investigate the coexistence of SR and CR.
Based on some exploration of the space of governing
parameters, we will end up with a conjecture on the
effective diffusive time scale of the core.
Actually, this effective diffusion time is
not well known for two reasons. The first one is
that the conductivity $\sigma$ of the
liquid Fe-Ni-Si alloy
at the pressure
of the Earth's core is hard to determine.
The most recent estimate \cite{STACEY}
is $\sigma=4.71 \times 10^5$ ($\Omega$ m)$^{-1}$.
For the Earth's core of radius $R=3480$ km
this would amount to a magnetic diffusion time
$T_d:=\mu_0 \sigma R^2=227$ kyr. The second reason
is the already mentioned possibility that  the ''molecular''
conductivity of the material could be reduced
due to the turbulent flow.
This so-called $\beta$ effect \cite{KRRA}
is very hard to estimate.
Actually, there have been claims \cite{REIGHARD} on the measurement
of a few percent $\beta$ effect in a turbulent
liquid sodium flow with magnetic Reynolds numbers $Rm$
up to 8,
but neither the Riga nor
the Karlsruhe
dynamo experiments have shown evidence of any
significant $\beta$ effect at much larger $Rm$ \cite{RMP}.
However, this negative result might not exclude
some $\beta$ effect in the geodynamo which works
at still higher $Rm$ and with a higher degree of
turbulence. Although a very strong reduction
(by a factor 10
or more) is safely excluded by
geomagnetic data, a reduction of $\sigma$
by a factor 2 or so is not completely forbidden and
will be a matter of consideration in this paper.

Our starting point is the well known 
induction equation \cite{KRRA} for evolution
of the magnetic
field ${\bm{B}}$
under the influence of a helical
turbulence parameter $\alpha$:
\begin{equation}
\frac{\partial {\bm{B}}}{\partial \tau} = {\bm \nabla} \times (\alpha {\bm{B}}) +
\frac{1}{\mu_0 \sigma} \Delta {\bm{B}} \; .
\end{equation}
After decomposing the magnetic field
into a poloidal and a toroidal parts according to
\begin{equation}
{\bm{B}}=-\nabla \times ({\bm{r}} \times \nabla S)-{\bm{r}} \times \nabla T \; ,
\end{equation}
the two defining scalars $S$ and $T$ are expanded in
spherical harmonics of degree $l$ and order $m$. Under the
assumption that
$\alpha$ is spherically symmetric (which is certainly a
grave simplification that does not apply to the Earth's
outer core) we arrive, for each degree $l$ and order $m$
separately,
at the following pair of partial differential equations:
\begin{eqnarray}
\frac{\partial s_l}{\partial \tau}&=& \frac{1}{r}\frac{\partial^2}
{\partial r^2}(r s_l)-\frac{l(l+1)}{r^2} s_l +\alpha(r,\tau) t_l \; ,\\
\frac{\partial t_l}{\partial \tau}&=&\frac{1}{r}\frac{\partial}{\partial r}
\left[ \frac{\partial}{\partial r}(r t_l)-\alpha(r,\tau)
\frac{\partial }{\partial r}(r s_l) \right] \; .\nonumber \\
&& - \frac{l(l+1)}{r^2}
[t_l-\alpha(r,\tau) s_l] \; .
\end{eqnarray}
Note that in
the non-dimensionalized equation system
(3-4) the radius $r$
is measured in units of $R$, the time $\tau$ in units of the
diffusion time $T_d$, and $\alpha$ in units of
$(\mu_0 \sigma R)^{-1}$.
Note further that the order $m$ of the spherical harmonics
does not show up in
equations (3-4) due to the
presupposed spherical symmetry of the problem.
The  boundary conditions are: $\partial s_l/\partial r
|_{r=1}+{(l+1)} s_l(1)=t_l(1)=0$.
In the following we will consider only the
dipole field with  $l = 1$.

\begin{figure}[ht]
\includegraphics[width=8cm]{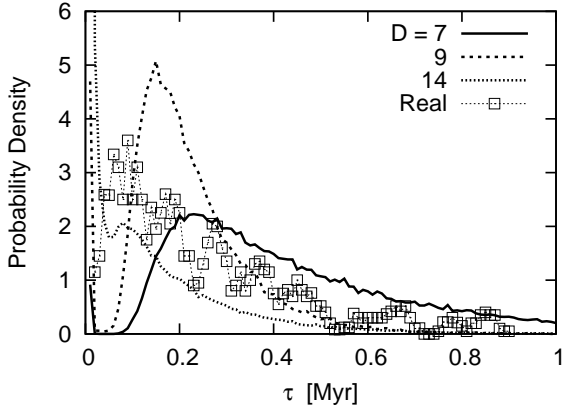}
\caption{RTD for $C=50$, $T_d=190$ kyr
and different
values of $D$ in the case without periodic driving,
compared to the ''real'' data from
paleomagnetic measurements which were
extracted from
\cite{CONSOL,LORITO}.}
\label{fig1}
\end{figure}

Beyond a critical intensity of $\alpha$, the
equation system (3-4) will acquire an exponentially
growing solution. In reality, however, the magnetic field
cannot grow  indefinitely, but will attenuate the
source of its own generation. We model this attenuation by
''quenching'' the kinematic $\alpha$ with the angle averaged
magnetic field energy which can be expressed in terms of $s(r)$
and $t(r)$. In addition, we assume that
$\alpha$ is also influenced by some noise which might be
considered as a shorthand for changing boundary conditions or
the neglected influence of higher magnetic field modes.
Taking into account quenching and multiplicative
noise together, we get
a time dependent $\alpha(r,\tau)$ in the form
\begin{eqnarray}
\alpha(r,\tau)&=&\frac{\alpha_{kin}(r)}{1 + E \left[ {\frac{2 s_{1}^{2}(r,\tau)}{r^2}+
\frac{1}{r^2}\left( \frac{\partial (r s_{1}(r,\tau))}
{\partial r} \right)^2 + t_{1}^{2}(r,\tau) } \right]   } \nonumber \\
&&+  \xi_1(\tau) + \xi_2(\tau)
 r^2 + \xi_3(\tau)  r^3+\xi_4(\tau) r^4 \; ,
\label{alpha}
\end{eqnarray}
with the noise correlation given by
\begin{eqnarray}
< \xi_i(\tau) \xi_j(\tau+\tau_1)>&=&D^2 (1-|\tau_1|/T_{c}) \nonumber \\
&&\times \Theta(1-|\tau_1|/T_{c}) \delta_{ij} \; ,
\end{eqnarray}
($\Theta$ is the Heaviside function).
In Eqs. (5,6), $\alpha_{kin}(r)$ is the kinematic $\alpha$ profile,
$D$ is
the noise intensity, $E$ is a
constant measuring the inverse mean field energy,
and $T_c$ is
a correlation time of the noise. As already mentioned
the diffusive time
scale $T_d$ is
approximately  200  kyr if we assume the molecular
conductivity of
the material
in the Earth's core. However, in the following we will consider
this value as variable.

Motivated by the earlier observation
that $\alpha$ profiles with
one sign change along the radius
can provide oscillatory solutions \cite{PRE,GIESECKE}, we choose for
the kinematic $\alpha$ profile in Eq. (5)
the Taylor expansion
\begin{eqnarray}
\alpha_{kin}(r)=1.914 \cdot C  \cdot ( \alpha_0 + \alpha_{1} r
+ \alpha_{2} r^{2} + \alpha_{3} r^{3} +
\alpha_{4} r^{4})
\end{eqnarray}
with $\alpha_0=1$,
$\alpha_1=\alpha_3=0$, $\alpha_2=-6$ and $\alpha_4=5$
(the factor 1.914 serves just
to normalize $C$ in order to make it comparable to the case
of constant $\alpha$).

\begin{figure}[ht]
\includegraphics[width=8cm]{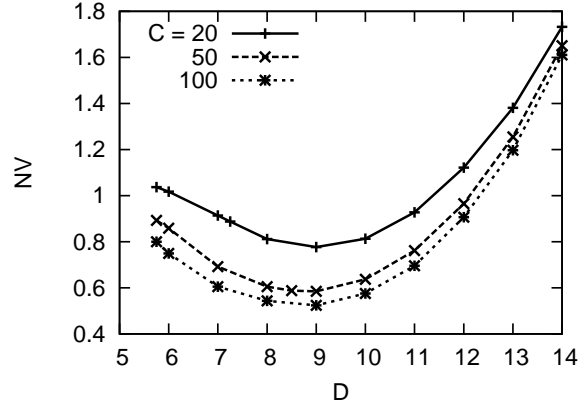}
\caption{Normalized variance ($NV$) of the RTD
for various $C$
in dependence on $D$ in the case without
periodic driving. The minimum at $D\sim 9$  is a typical
signature of CR.}
\label{fig2}
\end{figure}

For most of the
following simulations, the dynamo number is
set  to $C=50$ which is highly supercritical compared to the
critical value $C_{crit}=6.8$ which is specific to our
particular $\alpha(r)$ profile.
The equation system (3-5) is time-stepped by means of a standard
Adams-Bashforth method with radial grid spacing of 0.02 and
time step length of 2 $\times 10^{-5}$. The correlation time $T_c$
has been
set to 0.005 which would correspond
to 1 kyr in case that the
diffusion
time is set to 200 kyr.
The resulting time series show reversal sequences
quite similar to those of the geodynamo
\cite{STE2,STE3,GAFD}. As an important characteristic
of these sequences we will determine the
distribution of
residence times between two subsequent reversals which
we will abbreviate by RTD.

We start with the case without any periodic forcing
of $\alpha$.
Figure 1 shows the RTD
of the
dynamo model for different values of the
noise intensity $D$. The diffusion time was set to
$T_d=190$ kyr which is just the double of the
95 kyr period which was already  mentioned in \cite{LIU} and
which was also found in \cite{LORITO} as a good fit to
the paleomagnetic data.
For the sake of
comparison, we have included the curve resulting
from paleomagnetic
measurement as they were published in \cite{CONSOL,LORITO}
with its typical SR feature of several maxima
around multiples of
$95$ kyr.
All of
the numerically resulting curves exhibit a maximum,
followed by an
exponential
decrease of the probability. Starting with $D=7$, for which the
RTD has its maximum at $\sim 200$ kyr,
a trend of this maximum towards smaller values of $\tau$ is visible.
For
$D=9$ it is located at $\sim$150 kyr and for $D=14$ it is
at $\sim$100 kyr.
However, for those larger values of $D$ we
observe a much to steep
exponential decay, together with
an unphysical increase of reversal probability for
very small values of $\tau$. This is a first
indication that an assumed
diffusion time
of 190 kyr might be too large to explain the first
maximum at $\sim 95$ kyr observed in the paleomagnetic data.

To illustrate a typical feature of CR
we use the normalized variance $NV$ of the residence
time distribution, which is
defined as follows:
\begin{equation}
NV=\frac{\sqrt{Var(\tau)}}{<\tau>} \; .
\end{equation}
Figure 2 shows the  dependence of $NV$ on the noise strength $D$.
For all considered values of $C$,
there is a  minimum of $NV$ at $D\sim 9$ which is a
clear signature  for the occurrence of CR
\cite{PIKOVSKY}.  At this value of the noise intensity we get
an optimal amplification of
a residence time
corresponding to an inherent timescale of the dynamo.
The increase of $NV$ for values $D>9$ is explained by
the fact that for large $D$ the dynamo is
dominated by the noise and very
short residence times become more
probable as it was already visible in the curves for $D=9, 14$
in fig. 1.

\begin{figure}[ht]
\includegraphics[width=8cm]{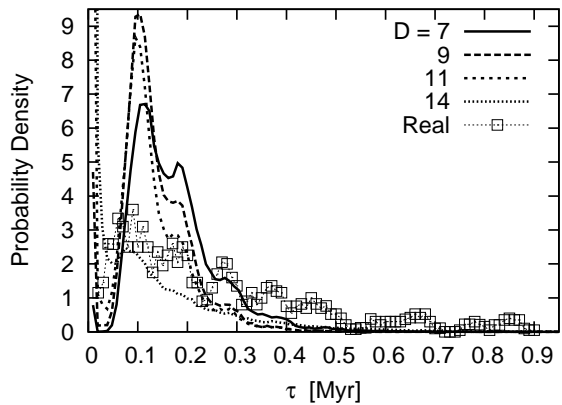}
\caption{RTD for $C=50$, $T_d=190$ kyr,
and different
values $D$ in the case with periodic driving with a
period of $T_{\Omega}=95$ kyr and
forcing strength $\epsilon=0.3$.}
\label{fig3}
\end{figure}
\begin{figure}[ht]
\includegraphics[width=8cm]{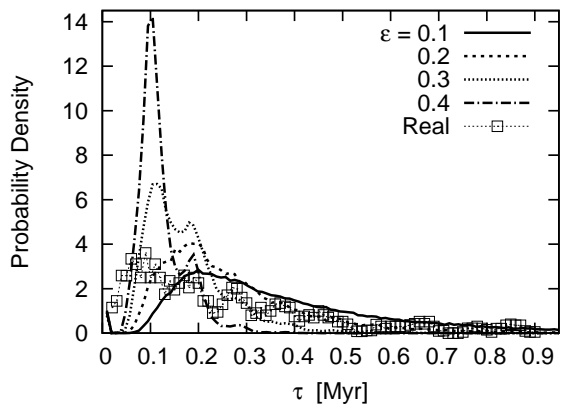}
\caption{RTD for $C=50$, $T_d=190$ kyr,
and different
values $\epsilon$ in the case with periodic driving with a
period of $T_{\Omega}=95$ kyr and
noise intensity $D=7$.}
\label{fig4}
\end{figure}

After having discussed the pure CR effect, we will now
examine the
additional occurrence of
SR. For this purpose,   a weak periodic
forcing of $\alpha$ is introduced. Here, ''weak'' means that
there are no
reversals in the absence of noise (note that a strong
periodic force would dominate the system completely
leading to a peak in the
residence time distribution located at the timescale of this
force). Although both additional and multiplicative noise
had been discussed in \cite{LORITO}, we focus here only on the
effects of multiplicative noise which we consider more physical
and which was also shown to fit better to the paleomagnetic
data
\cite{LORITO}. Actually, the periodic input is implemented as
an additional periodic variation of the helical turbulence
parameter
$\alpha$ with the period $T_{\Omega}$. Naively, one would
first apply the periodic forcing to the dynamo number $C$.
However, simulations have
shown that the dynamo does not react significantly
to such a homogeneous
periodic input. This insensitivity has to do with the
quenching effect which reduces the kinematic $\alpha$ to some
quenched $\alpha$ profile which makes the
growth rate of the dynamo close to zero. Contrary to such
a homogeneous periodic forcing of $\alpha$, a periodic change
of the {\it shape} of the $\alpha(r)$ profile has a more
pronounced effect.
As a simplest attempt, we have chosen to
add periodic forcing to the first coefficient $\alpha_0$ of the
Taylor expansion of $\alpha_{kin}(r)$
leaving all other coefficients unchanged:
\begin{equation}
\alpha_0(\tau) = 1 + \epsilon \cos( \frac{2 \pi}{T_{\Omega}} \cdot \tau)
\end{equation}
For the period $T_{\Omega}$ we chose always the dominant 95 kyr
contribution, although the total
effect of the Milankovich cycles is certainly much
more complicated \cite{LIU}. The parameter $\epsilon$
measures the relative strength of the periodic forcing.

\begin{figure}[ht]
\includegraphics[width=8cm]{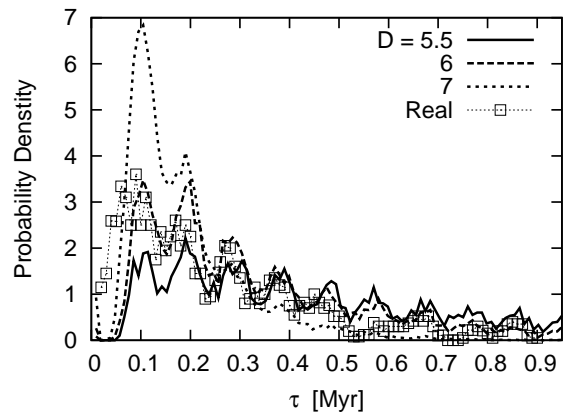}
\caption{RTD for $C=50$, $T_d=95$ kyr,
and different
values $D$ in the case with periodic driving with a period of $T_{\Omega}=95$ kyr and
forcing strength $\epsilon=0.1$.}
    \label{fig5}
\end{figure}
\begin{figure}[ht]
\includegraphics[width=8cm]{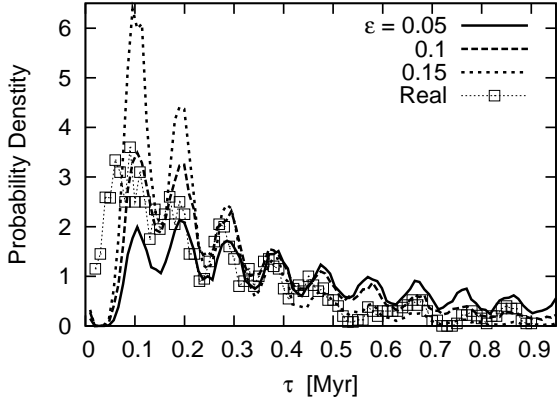}
\caption{RTD for $C=50$, $T_d=95$ kyr,
and different
values $\epsilon$ in the case with periodic driving with a period
of $T_{\Omega}=95$ kyr and
noise intensity $D=6$.}
    \label{fig6}
\end{figure}

The results for $C=50$, $T_d=190$ kyr, $T_{\Omega}=95$ kyr
are
shown in fig. 3 where we vary $D$ for fixed $\epsilon=0.3$,
 and in
fig. 4 where we vary $\epsilon$ for fixed $D=7$.
We see in fig. 3, that for D=7 the residence time distribution shows an
oscillatory behaviour with a maximum at $\tau=95$ kyr
and further
peaks at integer multiples of $T_{\Omega}$, which decay
exponentially.
Evidently, the oscillatory behaviour of the probability
becomes less pronounced for higher values of $D$. It is also
visible,
that for even higher values of $D$ the
probability at very small $\tau$
increases. The effect of increasing $\epsilon$ (for fixed $D=7$)
is quite similar to the effect of increasing $D$ (for fixed $\epsilon$).
However,
the increase of the probability density for very small $\tau$ does
not show up here.

\begin{figure}[ht]
\includegraphics[width=8cm]{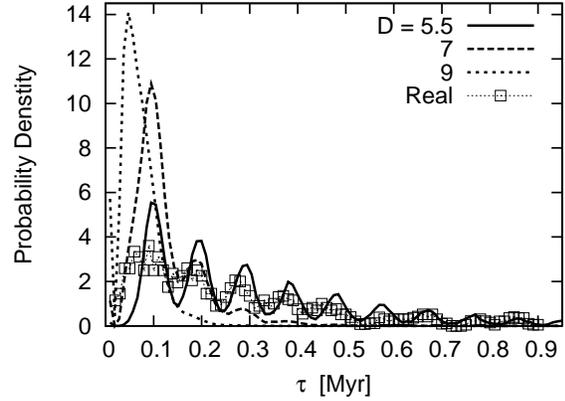}
\caption{RTD for $C=50$,
$T_d=63.3$ kyr and different
values $D$ in the case with periodic driving
with a period of $T_{\Omega}=95$ kyr and
forcing strength $\epsilon=0.1$.}
\label{fig7}
\end{figure}
\begin{figure}[ht]
\includegraphics[width=8cm]{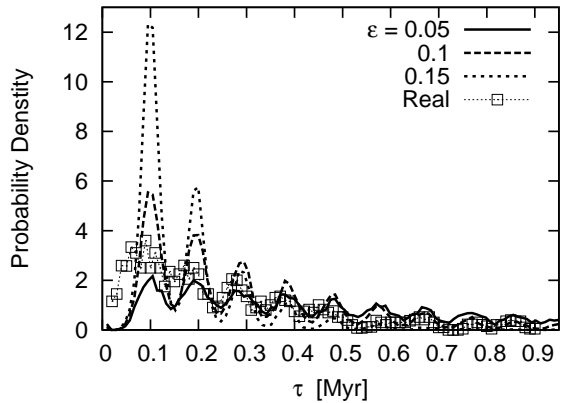}
\caption{RTD for $C=50$, $T_d=63.3$ kyr,
and different
values $\epsilon$ in the case with periodic driving with a period
of $T_{\Omega}=95$ kyr and
noise intensity $D=5.5$.}
    \label{fig8}
\end{figure}

Despite some similarity of the curves in figs. 3 and 4
with the paleomagnetic data,
with the given $T_{\Omega}=95$ kyr and $T_d=190$ kyr
it seems hard to
find parameters $\epsilon$ and $D$
which lead to a real good correspondence.
This has much to do with the fact that for
$T_d > T_{\Omega}$, which applies
to our example, the oscillatory
character of the probability distribution starts
only at $T_d$ as long as $D$ and $\epsilon$ are not
too large. This
behaviour
had already been demonstrated in \cite{CENT}.
To get the first peak at $\tau=T_{\Omega}=95$ kyr a rather
large value
of $\epsilon$ and/or $D$  is required which, in turn, leads to
a too steep exponential decrease of the probability function.

For this reason, we will test
in the following two additional values of
the diffusion time,
$T_d=95$ kyr $=T_{\Omega}$  (figs. 5 and 6) and   $T_d=63.3$ kyr $=2/3 \; T_{\Omega}$
(figs. 7 and 8).
For $T_d=95$ kyr, $\epsilon=0.1$, $D=7$ (fig. 5), but also
for $T_d=95$ kyr, $\epsilon=0.15$, $D=6$ (fig. 6)
we observe a quite good agreement with the
paleomagnetic data,
though the ratio between the first and
the second peak is
not exactly the same.

The version with $T_d=63.3$ kyr, $\epsilon=0.1$, $D=5.5$ (fig. 7)
looks also quite promising, although the
minima between
the probability peaks seem to be more pronounced than in
the paleomagnetic data. However, this argument might not be
so relevant
since in the paleomagnetic
data some stronger
''smearing out'' of the minima could
simply result from averaging over
a long time period (160 Myr), in which some parameters
must undergo some changes in order to explain
the long term
variation of the reversal rate (cp. \cite{MERRIL}, p. 184).

\begin{figure}[ht]
\includegraphics[width=8cm]{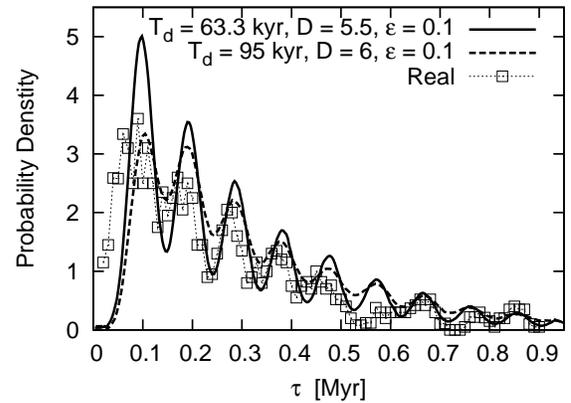}
\caption{RTD for the paleomagnetic data and two reasonable parameters for which
the moving box technique according to \cite{CONSOL}
has also been applied.}
\label{fig9}
\end{figure}

Another point is that in the derivation of the histogram of the
paleomagnetic data \cite{CONSOL}, the authors had used a moving box technique
which automatically
leads to some smoothing of the curve. If we do the same moving box technique
to our data, we arrive
at the curves shown in fig. 9, which are in reasonable correspondence
with the
paleomagnetic curve.

To summarize, we have shown that the
SR phenomenon which seems to lay at the root of the
95 kyr
periodicity of RTD of
the Earth's
magnetic field reversal
can coexist with the CR phenomenon
which is typical for our spherically symmetric $\alpha^2$
dynamo model.
The interesting question is now if the shape of the
residence time distribution can serve as a proxy
for determining the effective diffusion time of the
Earth's core and hence
for the reduction of the
conductivity due to the $\beta$ effect.
Although we had shown in \cite{STE3,GAFD} that a diffusion
time of
200 kyr is well compatible with the typical decay and
recovery times for individual reversals
(as long as some high super-criticality
is assumed) we see now that a diffusion time of
100 kyr or a bit below
seems more appropriate
when it comes to explain the stochastic resonance phenomenon.
Finally, this has to do with the effect that the first peak of the
stochastic resonance at $\sim$100 kyr is hardly explainable
with a diffusion time of $\sim$200 kyr, as long as the
strength of the periodic forcing and/or the noise are not
forbiddingly large.

Further work will focus on a systematic exploration of the
space of parameters $C$, $D$, $T_d$, and $\epsilon$ in order
to accommodate simultaneously
as much paleomagnetic constraints as possible.

\acknowledgments
This work was supported by Deutsche Forschungsgemeinschaft
in frame of SFB 609.
Stimulating discussions with Uwe G\"unther and  Andr\'e Giesecke
are gratefully acknowledged.

\end{document}